\documentclass[12pt]{article}
\usepackage{amsmath, amssymb}

\newtheorem{lemma}{Lemma}[section]

\newtheorem{theorem}{Theorem}[section]

\newtheorem{corollary}{Corollary}[section]

\newtheorem{proposition}{Proposition}[section]

\def \ddt {\frac{d}{dt}}

\def \DD  {{\cal D}}

\def \FF  {{\cal F}}

\def \NN  {{\cal N}}

\def \UU  {{\cal U}}

\def \WW  {{\cal W}}

\def \FFF {\mathbb{F}}

\def \HHH {\mathbb{H}}

\def \NNN {\mathbb{N}}

\def \RRR {\mathbb{R}}

\def \TR {\hbox{Tr}}

\title{Accuracy of the time-dependent Hartree-Fock approximation for
uncorrelated initial states}

\author{
Claude BARDOS%
\footnote{ Univ. Paris 7 \& Lab. J.-L. Lions, Bo\^\i te courrier
187, 75252 Paris cedex 05, France (bardos@math.jussieu.fr). },
Fran\c cois GOLSE%
\footnote{ Univ. Paris 7 \& Lab. J.-L. Lions, Bo\^\i te courrier
187, 75252 Paris cedex 05, France (golse@math.jussieu.fr). },
Alex D.\  GOTTLIEB %
\footnote{ Wolfgang Pauli Inst. c/o Inst.\ f.\ Mathematik, Univ.
Wien, Strudlhofg.\ 4, A--1090 Wien, Austria
(alex@alexgottlieb.com).
} \\
and
Norbert J.\ MAUSER%
\footnote{ Wolfgang Pauli Inst. c/o  Inst.\ f.\ Mathematik, Univ.
Wien, Strudlhofg.\ 4, A--1090 Wien, Austria
(mauser@courant.nyu.edu). } }

\date{ }

\begin{document}

\maketitle

\begin{abstract}
This article concerns the time-dependent Hartree-Fock (TDHF)
approximation of single-particle dynamics in systems of
interacting fermions.   We find that the TDHF approximation is
accurate when there are sufficiently many particles and the
initial many-particle state is any Gibbs equilibrium state for noninteracting
fermions (with Slater determinants as a special example).
Assuming a bounded two-particle interaction, we obtain
a bound on the error of the TDHF approximation, valid for short
times.  We further show that the error of the TDHF
approximation vanishes at all times in the mean field limit.
\end{abstract}

\section{Introduction}

The time-dependent Hartree-Fock (TDHF) equation is a nonlinear
Schr\"odinger equation designed to approximate the evolution of
an $n$-electron system.  The TDHF equation was first
written down by Dirac, both
as a system of $n$ coupled Schr\"odinger
equations for occupied orbitals, and as
an integro-differential equation for the ``density matrix," i.e., the integral kernel
$F(x,y,t)$ of the single-particle density operator \cite{Dirac,
Jamieson}.  In the latter form it reads
\begin{eqnarray}
     i \frac{d}{dt} F(x,y,t) & = & -\tfrac{1}{2}(\Delta_x -
     \Delta_y)F(x,y,t) \ + \ (V_{ext}(x) -
     V_{ext}(y))F(x,y,t)  \nonumber   \\
         & + &   \int_{\RRR^3}
         \big[ V(|x-z|) - V(|y-z|) \big]
          F(z,z,t)dz F(x,y,t)
   \nonumber   \\
     & - &
   \int_{\RRR^3} \big[
             V(|x-z|) -  V(|y-z|)
             \big]
             F(x,z,t)F(z,y,t) dz     \label{exchange}
\end{eqnarray}
in atomic units, where $V_{ext}$ is the external potential energy and
$V(r)=1/r$ is the Coulomb interaction potential.
The the last term on the right-hand side of (\ref{exchange}) is the
``exchange" term.

The Coulomb potential, however, is not amenable to the
techniques of this article because it is unbounded (this case being dealt with in \cite{Manuscript}).
In this article, we consider interaction potentials given by a
bounded function $V(|x-y|)$, or, more generally,
any bounded, symmetric, two-body operator $V$ (not necessarily a multiplication operator).
The one-particle energy operator will be denoted by $L$, the
interaction energy operator for a single pair of particles will be denoted by $V$,
and the total energy operator for a system of particles will be
the sum of all single-particle energies and all pair energies.
  Although the number of particles does not change
under the dynamics just described, we prefer to formulate the
dynamics on a fermion Fock space so that we may consider initial
states of indeterminate particle number. We are going to show that quasifree
initial states enhance the accuracy of the TDHF approximation.

  Let $L$ be a
self-adjoint operator on a Hilbert space $\HHH$, and let $V$ be a bounded
Hermitian operator on $\HHH \otimes \HHH$ that commutes with the
transposition operator $U$ defined by $U(x\otimes y)=y\otimes x$.
We are going to discuss the dynamics whose Hamiltonian $H$ on the
fermion Fock space $\FFF_{\HHH}$ is written as
\begin{equation}
\label{Ham}
        H \ = \ \sum_{i,j} \langle j| L| i \rangle a^{\dagger}_j a_i
        \ + \ \sum_{i,j,k,l} \langle k l | V | i j \rangle
        a^{\dagger}_k
        a^{\dagger}_l a_j a_i
\end{equation}
in second quantized form.
    We will analyze the solutions of the von
Neumann equation
\begin{eqnarray}
          i  \frac{d}{dt} D(t) & = & [H,D(t)]   \nonumber \\
                D(0) & = & D_0
          \label{vN},
\end{eqnarray}
which is the evolution equation for the density operator on Fock space in the
  Schr\"odinger picture of quantum dynamics (in units of time and
energy for which $\hbar=1$).

  We will see that (\ref{vN}) leads to the following equation for
the single-particle (number) density operator $\NN_1(t)$:
\begin{eqnarray}
          i  \frac{d}{dt} \NN_1(t) & = & [L,\NN_1(t)] \ + \ [V, \NN_2(t) ]_{:1}
      \nonumber  \\
                               \NN_1(0) & = & \NN_1(D_0),
      \label{vN1}
\end{eqnarray}
where $[V, \NN_2(t) ]_{:1}$ denotes the partial trace of the
two-particle operator $[V, \NN_2(t) ]$.  Equation (\ref{vN1}) for
$\NN_1(t)$ is not ``closed" since its right hand side involves the
two-particle density operator $\NN_2(t)$.
  The TDHF approximation to $\NN_1(t)$ is the solution of the initial
value problem
\begin{eqnarray}
   i \ddt F(t) & = & [L,F(t)] \ + \ [V,\ F(t)^{\otimes 2}2A_2]_{:1}
\nonumber  \\
   F(0) & = & \NN_1(D(0))
   \label{TDHF}
\end{eqnarray}
where $A_2$ is the orthogonal projector of $\HHH\otimes \HHH$ onto
the subspace of antisymmetric vectors.  The existence and
uniqueness of solutions of (\ref{TDHF}) were established in
\cite{Bove1} for the case where $V$ is a bounded operator, and in
\cite{Chadam, Bove2} for the case where $V$ is a Coulombic
interaction.

The TDHF equation (\ref{TDHF}) is obtained by closing the
single-particle equation (\ref{vN1}) with the Ansatz
\begin{equation}
\label{Ansatz}
         \NN_2 = (\NN_1\otimes \NN_1)2A_2
\end{equation}
at all times.  The relation (\ref{Ansatz}) holds for pure states
corresponding to Slater determinants, and also for Gibbs
densities.  However, even supposing that $\NN_2(0)$ satisfies
(\ref{Ansatz}), the interaction $V$ is likely to introduce
``correlations" in $\NN_2(t)$, that is, departures from
(\ref{Ansatz}), and ignoring those correlations in the TDHF
equation requires justification.

We are going to prove that the absence of correlations is
self-perpetuating in the mean field limit.
Theorem~\ref{mean-field-theorem} states that if $\NN_2(0)$
satisfies (\ref{Ansatz}) then $\NN_2(t)$ asymptotically satisfies
(\ref{Ansatz}) as the number of particles $n$ tends to infinity
and the interaction strength is scaled as $1/n$.
In this scaling, the force exerted on each individual particle by
the $n-1$ other particles is of $O(1)$ as $n \to \infty$.  This was
called the ``mean-field scaling''
by H. Spohn in his fundamental review paper \cite{Spohn}, where he
derives the time-dependent Hartree equation.
In the appendix we prove an important special case of Theorem~5.7 of \cite{Spohn} as a corollary of our
Theorem~\ref{mean-field-theorem}.

Technically, we rely on the trace norm
approach used in \cite{Spohn} to derive the Hartree equation.
We first published our derivation of the TDHF equation in the mean
field limit in \cite{BGGM1} for initial
states of fixed particle number such as Slater determinants. The main
advance of this article
is that the initial states need not be Slater determinants; the
TDHF approximation should work equally well (or badly) for any
quasifree initial state. Also, in this article we are not only
interested in the mean field limit, and we derive the error bound
of Theorem~\ref{short-times} for the unscaled problem.

Throughout this paper, $n$ denotes a fixed number of particles, $N$ denotes the
the number operator on the Fock space over $\HHH$, and $\NN_1$ denotes the number density operator on $\HHH$.


\section{Definitions and notation}
\label{notation}

  Consider a quantum particle whose Hilbert space is $\HHH$, i.e.,
  a particle which, in isolation, would constitute
a system whose (pure) quantum states are represented by the
rank-one orthogonal projectors on some Hilbert space $\HHH$. The
set of quantum states available to a system of $n$ particles of
this kind is determined by their ``statistics," i.e., whether the
particles are fermions, bosons, or distinguishable.  If the
particles are fermions, a (pure) state of a system of $n$ of them
is represented by a rank-one projectors onto vectors in the
antisymmetric subspace $\HHH^{(n)}$ of the tensor power space
$\HHH^{\otimes n}$.  If the number of fermions in the system is
not fixed, the appropriate Hilbert space is the fermion Fock space
\begin{equation}
\label{FockSpace}
      \FFF_{\HHH} \ =
\   \HHH^{(0)} \oplus \HHH^{(1)} \oplus \HHH^{(2)} \oplus \HHH^{(3)}
\oplus \cdots \qquad .
\end{equation}
The possibility of a zero-particle state is accommodated by
$\HHH^{(0)}$, a one-dimensional space spanned by the vacuum
vector.  We denote the number operator on $\FFF_{\HHH}$ by $N$.

Let $\Pi_n$ denote the group of permutations of
$\{1,2,\ldots,n\}$.  For each $\pi \in \Pi_n$, a unitary operator
$U_{\pi}$ on $\HHH^{\otimes n}$ may be defined by extending
\[
         U_{\pi}(x_1 \otimes x_2 \otimes \cdots \otimes x_n) \ = \
         x_{\pi^{-1}(1)} \otimes x_{\pi^{-1}(2)} \otimes \cdots
\otimes x_{\pi^{-1}(n)}
\]
to all of $\HHH^{\otimes n}$.  The operator
\begin{equation}
\label{An}
     A_n \ = \ \frac{1}{n!}\sum_{\pi \in \Pi_n} \hbox{sgn}(\pi)
     U_{\pi}
\end{equation}
is the orthogonal projector with range $\HHH^{(n)}$.  If
$x_1,x_2,\ldots,x_n$ is an orthonormal system in a Hilbert space
$\HHH$, then the vector
\begin{equation}
\label{SlaterD}
        \sqrt{n!} \ A_n(x_1 \otimes x_2 \otimes \cdots \otimes
      x_n),
\end{equation}
is a unit vector in $\HHH^{(n)}$, called a {\it Slater
determinant}.

We will consider many-particle states that are represented by
density operators on $\FFF_{\HHH}$.  We will only consider density
operators that commute with $N$ and such that $N^mD$ is trace
class for all $m \in \NNN$.  For such densities $D$, one can
define the reduced density operators $\NN_m(D)$ of all orders $m$:

If $T$ is a trace class operator on $\HHH^{(n)}$, and $m \le n$,
we will use the subscript notation $T_{:m}$ to denote the partial
trace of $T$ of order $m$, a trace class operator on $\HHH^{(m)}$.
(This operator is defined unambiguously thanks to the symmetry of
$T$ considered as an operator on $\HHH^{\otimes n}$.)  If $T$ is a
density operator on $\HHH^{(n)}$, the operator $T_{:m}$ is known
as the {\it m-particle reduced density operator} \cite{TerHaar}
and it is used to determine the expected values of the
$m$-particle observables.
   Let $D$ be a density operator on $\FFF_{\HHH}$ that commutes with $N$.
Then
\begin{equation}
\label{directsum}
     D \ = \ \bigoplus_{n=0}^{\infty} D_n
\end{equation}
where each $D_n$ is a nonnegative trace class operator on
$\HHH^{(n)}$.  Assuming that
\begin{equation}
\label{m-moment}
    \TR(N^mD) \ = \  \sum_{n=0}^{\infty} n^m \TR(D_n) \ < \ \infty,
\end{equation}
we may define the $m^{th}$ order reduced density operator
\begin{equation}
\label{sumofmarginals}
      \NN_m(D) \ = \  \sum_{n=m}^{\infty}\frac{n!}{(n-m)!} D_{n:m},
\end{equation}
where $D_{n:m} = (D_n)_{:m}$ is the partial trace of $D_n$.  The
reduced density $\NN_m$ serves to describe the $m$-particle
correlations in a system of many particles (see Section 6.3.3 of
\cite{Bratteli}).


\section{Dynamics and the BBGKY hierarchy} \label{carefully}

We now define the dynamics (\ref{vN}) and the reduced dynamics
(\ref{vN1}).

Let $H$ be a self-adjoint operator on $\HHH$, and let $V$ be a
bounded Hermitian operator on $\HHH\otimes \HHH$ that commutes
with the transposition operator $U(x\otimes y)=y\otimes x$. For $1
\le j \le n$, let $L_j$ denote the operator
\[
       \stackrel{j-1  \ \  times}{I \otimes \cdots \otimes I}
     \otimes L \otimes
     \stackrel{n-j  \ \  times}{I \otimes \cdots \otimes I}
\]
on $\HHH^{\otimes n}$ (the value of $n\ge j$ is not explicit in
the notation $L_j$ but it will always be clear from context). For
$1 \le i < j \le n$, let $U_{(ij)}$ denote the permutation
operator on $\HHH^{\otimes n}$ that transposes the $i^{th}$ and
$j^{th}$ factors of any simple tensor $x_1 \otimes \cdots \otimes
x_n$, and let
\[
     V_{ij}\ = \  U_{(1i)}U_{(2j)} \left( V \otimes I^{\otimes n-2}
     \right) U_{(2j)}U_{(1i)}
\]
(again, the domain $\HHH^{\otimes n}$ of $V_{ij}$ will always be
clear from context).
  For each $n$, define the operators
\begin{eqnarray*}
    L^{(n)} & = & \sum\limits_{j=1}^n L_j  \\
    H^{(n)} & = & L^{(n)} \ + \ \sum\limits_{1\le i < j \le n} V_{ij}
\end{eqnarray*}
on $\HHH^{(n)}$ (these operators are defined on all of
$\HHH^{\otimes n}$ but we are only considering their restrictions
to the invariant subspace $\HHH^{(n)}$). The Hamiltonian operator
$H$, which we had formally represented above by (\ref{Ham}), is
the direct sum $H = \bigoplus H^{(n)}$ defined on the domain
\[
     \DD(H) \ = \ \Big\{
     x = \oplus x_n \in \FFF_{\HHH}: \ \sum_n \big\| H^{(n)}x_n \big\|^2 \ < \
     \infty
     \Big\}.
\]
This operator is closed and self-adjoint (see Section 6.3.1 of
\cite{Bratteli}) and $-iH$ is the generator of the strongly
continuous group
\[
     W_t \ = \  \bigoplus_{n=1}^{\infty} W^{(n)}_t
\]
of unitary operators on $\FFF_{\HHH}$, where $W^{(n)}_t =
\exp\big(-itH^{(n)}\big)$.  The dynamics corresponding to
(\ref{vN}) are given by the group
\begin{equation}
      \WW_t(D) \ = \    W_t D W_{-t}
\label{dynamics}
\end{equation}
of isometries of the space of Hermitian trace class operators on
$\FFF_{\HHH}$. (See Proposition 3.4 of \cite{Bove1} for a proof
that groups of isometries defined in this way are strongly
continuous.)  It is convenient to have some notation for the free
part of the dynamics, so we define
\begin{equation} \label{noninteracting}
      \UU^{(n)}_t(T) \ = \    U^{(n)}_t T U^{(n)}_{-t}
\end{equation}
with $  U^{(n)}_t =  \exp\big(-itL^{(n)}\big) $.

The dynamical equation for the $m^{th}$ order reduced density
$\NN_m$ can be derived from (\ref{dynamics}) if the density $D$
satisfies the moment condition (\ref{m-moment}).  The details of
the derivation are provided in \cite{Preprint}.
\begin{proposition}
\label{BBGKY}
  Let $\UU_t$ and $\WW_t$ be as defined in
(\ref{noninteracting}) and (\ref{dynamics}).

Suppose that $D$ is a density operator on $\FFF_{\HHH}$ of the
form $D=\oplus D_n$ such that (\ref{m-moment}) holds for some $m
\in \NNN$. Let $\NN_m(t)$ denote $\NN_m(\WW_t(D))$.  Then
$\NN_m(t)$ satisfies
\[
      \NN_m(t) \ = \
       \UU^{(m)}_t \NN_m(0) \ - \ i \int_0^t \UU^{(m)}_{t-s}
               \sum_{1 \le i< j \le m}\left[ V_{ij}, \NN_m(s) \right]ds
         \ - \
          i \int_0^t \UU^{(m)}_{t-s}
          \sum_{j=1}^m \left[V_{j,m+1}, \NN_{m+1}(s)
          \right]_{:m}ds\ .
\]
\end{proposition}
These equations for the reduced density operators are known as the
BBGKY hierarchy.  The first equation of the hierarchy is equation
(\ref{vN1}) in integral form.


\section{The TDHF hierarchy}
\label{hartreehierarchy}

The existence and uniqueness of mild solutions of the TDHF
equation (\ref{TDHF}) is established in \cite{Bove1}.   There it
is shown that the integral equation
\begin{equation}
  F(t) \ = \  \UU^{(1)}_t F(0) \ - \ i \int_0^t \UU^{(1)}_{t-s}
             \left[V, \ F(s)^{\otimes 2} 2A_2\right]_{:1}ds
  \label{TDHF-mild}
\end{equation}
has a unique solution $F(t)$ for any Hermitian trace class
operator $F(0)$.   Define $\FF_1(t) = F(t)$ and, for $m > 1$,
define
\begin{equation}
  \FF_m(t) \ = \  F(t)^{\otimes m}m! A_m.
\label{FFm}
\end{equation}

We proceed to derive equations for the $\FF_m(t)$ from
(\ref{TDHF-mild}).

First, set $G(t) = \UU^{(1)}_{-t} F(t)$, so that
\[
     G(t) \ = \ F(0) \ - \ i \int_0^t \UU^{(1)}_{-s}
             \left[V, \ F(s)^{\otimes 2} 2A_2\right]_{:1}ds.
\]
Now apply the product rule (in integral form) to $G(t)^{\otimes
m}$:
\begin{eqnarray*}
     G(t)^{\otimes m} & = &
        F(0)^{\otimes m} \ - \
        i \sum_{j=1}^m \int_0^t  G(s)^{\otimes j-1} \otimes \UU^{(1)}_{-s}
             \left[V, \ F(s)^{\otimes 2} 2A_2\right]_{:1} \otimes
G(s)^{\otimes n-j}ds
           \\
       & = &
       F(0)^{\otimes m} \ - \
       i \int_0^t \UU^{(m)}_{-s} \sum_{j=1}^m
       \Big[V_{j,m+1}, \ F(s)^{\otimes m+1}
\big(I-U_{(j,m+1)}\big)\Big]_{:m} ds .
\end{eqnarray*}
Apply $\UU^{(m)}_t$ to both sides of the preceding equation to obtain
\[
     F(t)^{\otimes m} \ = \  \UU^{(m)}_t F(0)^{\otimes m} \ - \
     i \int_0^t \UU^{(m)}_{t-s} \sum_{j=1}^m
     \Big[V_{j,m+1}, \ F(s)^{\otimes m+1} \big(I-U_{(j,m+1)}\big)\Big]_{:m} ds .
\]
Multiply both sides of the last equation by $m!A_m$ on the left,
noting that $\UU^{(m)}_s(X)A_m = \UU^{(m)}_s(XA_m)$ and that $A_m$
commutes with $\sum V_{j,m+1}$:
\[
     \FF_m(t) \ = \ \UU^{(m)}_t \FF_m(0) \ - \
     i \int_0^t \UU^{(m)}_{t-s} \sum_{j=1}^m
     \Big[V_{j,m+1}, \ F(s)^{\otimes m+1} \big(I-U_{(j,m+1)}\big)m!A_m
\Big]_{:m} ds .
\]
Since $(m+1)!A_{m+1} = \big(I - U_{(1,m+1)} \  \cdots \ -
U_{(m,m+1)}\big)m!A_m$, the last equation may be rewritten
\begin{eqnarray}
     \FF_m(t) & = &
      \UU^{(m)}_t \FF_m(0) \ - \
     i \int_0^t \UU^{(m)}_{t-s} \sum_{j=1}^m
     \left[V_{j,m+1}, \ \FF_{m+1}(s)  \right]_{:m} ds  \nonumber \\
     & + &
       \sum_{1 \le j \ne k \le m} i \int_0^t \UU^{(m)}_{t-s}
     \Big[V_{j,m+1}, \  F(s)^{\otimes m+1} U_{(k,m+1)}m!A_m \Big]_{:m} ds\  .
\label{FFm(t)}
\end{eqnarray}
We call these equations for the $\FF_m(t)$ the {\it TDHF
hierarchy}.


\section{Estimates} \label{estimates}

In this section we have collected some estimates used in the
proofs of Theorems~\ref{short-times} and \ref{mean-field-theorem}.
The first two of the following propositions are stated here
without proof, for the facts are known, and the reader may find
proofs of them in \cite{Preprint}.  We use the notation $\|\cdot\|$ for the operator norm and
$\|\cdot\|_1$ for the trace norm.

\begin{proposition}
\label{N1} If $D$ is a density operator on $\FFF_{\HHH}$ that
commutes with $N$ and such that $\TR(ND)<\infty$, then the
operator norm of $\NN_1(D)$ is not greater than $1$.
\end{proposition}

\begin{proposition}
\label{Prop1.5} If $T$ is a Hermitian trace class operator then
\begin{equation}
\label{proveMe}
        \big\| T^{\otimes n}n!A_n  \big\|_1 \ \le     \big\| T
        \big\|_1^n.
\end{equation}
\end{proposition}

\begin{proposition}
\label{TDHFbound}

The trace norm of the last term in (\ref{FFm(t)}) does not exceed
\[
        2m(m-1)\|V\|\ \|F(0)\| \ \big\|F(0)\big\|_1^m t \ .
\]
\end{proposition}

\noindent {\bf Proof:} \qquad  The trace norm of the last term in
(\ref{FFm(t)}) is bounded by
\begin{equation}
\label{pp-important-estimate}
        2m(m-1) \int_0^t \Big\| \left\{ V_{m-1,m+1} U_{(m,m+1)}
        \big(\FF_m(s)\otimes F(s)\big)\right\}_{:m} \Big\|_1 ds
\end{equation}
thanks to the symmetry of $\FF_m$.  It can be verified that
\[
\left( V_{m-1,m+1}U_{(m,m+1)} (\FF_m(s)\otimes F(s)) \right)_{:m}
\ = \ \big(I^{\otimes m-1}\otimes F(s) \big)V_{m-1,m} \FF_m(s),
\]
whence (\ref{pp-important-estimate}) is bounded by
\[
        2m(m-1) \int_0^t
        \|V\| \ \|F(s)\| \ \big\|\FF_m(s)\big\|_1 ds.
\]
Now $\|\FF_m(s)\|_1 \le \| F(s) \|_1^m$ by
Proposition~\ref{Prop1.5} since $\FF_m=F^{\otimes m}m!A_m$.
Furthermore, $\|F(s)\|_1 = \|F(0)\|_1$ and $\|F(s)\| = \|F(0)\|$
for all $s>0$ by Proposition~4.3 of \cite{Bove1}.  Thus we arrive
at the bound stated in the Proposition. \hfill $\square$


\section{Accuracy of the TDHF approximation}
\label{accuracy}

In this section we will compare the single-particle density
operator $\NN_1(t)$ to its approximation by the solution of the
TDHF equation
\begin{eqnarray}
  F(t) & = &
   \UU^{(1)}_t F(0) \ - \ i \int_0^t \UU^{(1)}_{t-s}
             \left[V, \ F(s)^{\otimes 2} 2A_2\right]_{:1}ds \nonumber \\
  F(0) & = & \NN_1(0).
  \label{TDHF-repeat}
\end{eqnarray}
We can control the distance between $\NN_1(t)$ and $F(t)$ in trace
norm when the initial many-particle state is gauge-invariant
quasifree. Theorems~\ref{short-times} and \ref{mean-field-theorem}
show how the accuracy of the TDHF approximation is enhanced when
the initial state involves many uncorrelated fermions.

%
%
We consider the class of initial states that have density operators $D$
of the form (\ref{directsum}) such that $\TR(ND)<\infty$ and
\begin{equation}
\label{exactClosure}
       \NN_n(D) \ = \  \NN_1(D)^{\otimes n}n!A_n
\end{equation}
for all $n \in \NNN$. These relations characterize the gauge-invariant
quasifree states of the CAR algebra having a trace class single-particle
reduced density operator $\NN_1$ \cite{Bratteli}.

There are two important examples of such initial states: Slater
densities and Gibbs grand canonical equilibrium states. ``Slater densities"
are those of the form
\[
     D \ = \   \stackrel{n-1  \ \  times}{{\bf 0} \oplus \cdots \oplus {\bf 0}}
     \big( |\psi\rangle\langle \psi| \big)
     \oplus {\bf 0} \oplus \cdots\ ,
\]
where $\psi$ is an $n$-particle Slater determinant (\ref{SlaterD}). Gibbs
equilibrium states are obtained as follows: let $H$ be the
single-particle Hamiltonian and $d\Gamma(H)$ its functorial extension to
the fermion Fock space viewed as the exterior algebra of $\HHH$. In
other words,
\[
d\Gamma(H)=\bigoplus_{n\ge 0}d\Gamma_n(H)
\]
where $d\Gamma_n(H)$ is the restriction to $\HHH^{(n)}$ of the operator
\[
\sum_{j=1}^n\stackrel{j-1 \ \ times}{I\otimes\ldots\otimes I}\otimes
     H\otimes\stackrel{n-j \ \ times}{I\otimes\ldots\otimes I}
\]
Whenever $\beta>0$ is such that $e^{-\beta H}$ is trace-class on $\HHH$,
the Gibbs equilibrium state at inverse temperature $\beta$ with chemical
potential $\mu$ is defined by the density operator proportional to
$\exp(-\beta d\Gamma(H-\mu I))$ (see Proposition 5.2.22 of \cite{Bratteli}
for more details on Gibbs states).


\begin{theorem}
\label{short-times}
  Let $D$ be the density operator on $\FFF_{\HHH}$ of a
  gauge-invariant quasifree state with finite expected particle
number, i.e., with $\TR(ND)<\infty$.
Let $\NN_1(t)$ denote $\NN_1(\WW_t(D))$, where $\WW_t$ is the
dynamics with two-particle interactions defined in
(\ref{dynamics}).

Let $F(t)$ be the
solution of the TDHF equation (\ref{TDHF-repeat}).

Let $\tau$ denote $(2\|V\|\| \NN_1 \|_1)^{-1}$. Then
\begin{equation}
\label{short-time-error}
          \big\| \NN_1(t) - F(t) \big\|_1
               \ \le \
               \frac{3}{2}
           \left( \frac{t}{\tau -  t}\right)^2
\end{equation}
for $t < \tau$.
\end{theorem}
The proof of this theorem is postponed until the end of this Section.

Unfortunately, the bound (\ref{short-time-error}) on the error of
the TDHF approximation is valid only when $t < \tau$ and we have
no explicit bounds for larger $t$.  In principle, the estimate of
Theorem~\ref{short-times} could be used to establish that any
effect observed in a TDHF simulation before the critical time
$\tau$ reflects a true effect of the interaction.  However, in the
numerical tests we have conducted so far, very little appears to
happen before the critical time $\tau$, and we fear that the
estimate of Theorem~\ref{short-times} might not prove generally
useful.
  Nonetheless, Theorem~\ref{short-times} does show that the error of the
TDHF approximation is less than one might expect when the initial
condition is an uncorrelated many-fermion state, for even at short
times $t < \tau$ the left-hand side of (\ref{short-time-error}) is
proportional to $\| F(0) \|_1=\|\NN_1(D(0))\|_1$ {\it prima facie}, not bounded
independently of $\|\NN_1\|_1$.

  The improvement in accuracy of the TDHF approximation when the initial
state is uncorrelated is even more evident in the ``mean
field scaling." It is in this spirit that we are about to
introduce a coupling constant $\lambda$ into the interaction term
of the many-particle system and consider a scalings where
$\lambda$ times the average particle number tends to $0$ or
remains bounded.   We do not discuss the physical significance
of such scalings; we consider them only so that we may more easily express how the accuracy of the
TDHF equation is affected by uncorrelated initial data.

  For each value of the parameter $\lambda > 0$, consider
the Hamiltonian
\begin{equation}
\label{LambdaHam}
        H_{\lambda} \ = \ \sum_{i,j} \langle j| L| i \rangle a^{\dagger}_j a_i
        \ + \ \lambda\sum_{i,j,k,l} \langle k l | V | i j \rangle
        a^{\dagger}_k
        a^{\dagger}_l a_j a_i.
\end{equation}
If the initial density operator represents a gauge-invariant
quasifree state with finite expected particle number, then
Proposition~\ref{BBGKY} implies that all reduced number density
operators exist and satisfy
\begin{eqnarray}
      \NN_{\lambda m}(t) & = &
       \UU^{(m)}_t \NN_{\lambda m}(0) \ - \
               \lambda i \int_0^t \UU^{(m)}_{t-s}
               \Big(
               \sum_{1 \le i< j \le m}
               \left[ V_{ij}, \NN_{\lambda m}(s) \right] \ + \
                  \sum_{j=1}^m \left[V_{j,m+1},
(\NN_{\lambda})_{m+1}(s) \right]_{:m}
              \Big)
              ds
              \nonumber \\
       \NN_{\lambda m}(0) & = & \NN_m(D_{\lambda}(0)) .
\label{LambdaNNm}
\end{eqnarray}
The TDHF equation corresponding to (\ref{LambdaHam}) is
\begin{eqnarray}
  F_{\lambda}(t) & = &  \UU^{(1)}_t F_{\lambda}(0)
             \ - \ \lambda i \int_0^t \UU^{(1)}_{t-s}
             \left[V, \ F_{\lambda}(s)^{\otimes 2} 2A_2\right]_{:1}ds
         \nonumber \\
  F_{\lambda}(0) & = & \NN_{\lambda 1}(0).
  \label{TDHF-lambda}
\end{eqnarray}

\begin{theorem}
\label{mean-field-theorem}

Let $\left\{D_{\lambda}\right\}_{\lambda >0}$ be a family of
density operators on $\FFF_{\HHH}$ that represent gauge-invariant
quasifree states with finite expected particle number.   Let
$\NN_{\lambda m}(t)$ be the solution of (\ref{LambdaNNm}) with
initial condition $N_{\lambda m}(0) = N_m(D_{\lambda})$ and let
$F_{\lambda}(t)$ be the solution of the TDHF equation
(\ref{TDHF-lambda}).  Let $\FF_{\lambda m}(t) =
F_{\lambda}(t)^{\otimes m}m! A_m$.
\begin{trivlist}
\item{\rm{1})\quad}
If $ \ \lim\limits_{\lambda\rightarrow 0} \lambda\|\NN_{\lambda 1
}(0)\|_1 = 0$ then, for each fixed $t>0$ and $m \in \NNN$,
\[
       \big\| \NN_{\lambda m}(t) - \FF_{\lambda m}(t)
       \big\|_1 \big/ \|\NN_{\lambda 1}\|_1^{m}
       \ = \ O(\lambda).
\]
\item{\rm{2})\quad} If $ \ \limsup\limits_{\lambda\rightarrow 0}
\lambda\|\NN_{\lambda 1}(0)\|_1 < \infty$ then
\[
       \lim\limits_{\lambda \rightarrow 0}
       \big\| \NN_{\lambda m}(t) - \FF_{\lambda m}(t)
       \big\|_1 \big/ \|\NN_{\lambda 1}\|_1^{m}
       \ = \ 0
\]
for all $t>0$ and all $m \in \NNN$.
\end{trivlist}
\end{theorem}

\noindent {\bf Remark on the persistence of the interaction in the mean field limit}

Although we have assumed only that the interaction $V$ is a bounded and symmetric two-particle operator, we are mainly interested in the case where it is a bounded multiplication operator.  In this case one may bound the exchange term in the TDHF equation as follows.  The exchange term  in (\ref{exchange}) is the sum of two products:
the product of the integral operators with kernels $V(|x-z|)
F(x,z)$ and $F(z,y)$, and the product of the integral
operators with kernels $F(x,z)$ and $V(|y-z|) F(z,y)$. The
trace norm of each of these products is bounded by the product of
the Hilbert-Schmidt norms of its factors. Since the operator norm of
a fermionic single-particle operator is less than or equal to $1$
(viz. Proposition~\ref{N1})
\[
       \| F \|_{HS} \ = \ \sqrt{\hbox{Tr}(F^2)}
       \ \le \ \sqrt{\| F \|_1 \| F \|} \ \le \ \sqrt{\| F \|_1}
\]
and it follows that the trace norm of the exchange term is bounded
by $2  \| F \|_1 \| V \|_{L^{\infty}} $. Thus the contribution of the
exchange term is not much larger than the error of the TDHF
approximation itself!  To see this more clearly, let us return to
Statement~1 of Theorem~\ref{mean-field-theorem}. Squeezing the most
we can out of its proof informs us that
\[
       \big\| \NN_{\lambda 1}(t) - F_{\lambda}(t)
       \big\|_1
       \ = \ O\big(\lambda^2 \| F_{\lambda} \|^2_1\big)
\]
if $ \ \lim\limits_{\lambda\rightarrow 0} \lambda\| F_{\lambda}
\|_1 = 0$. (It is due to the fact that the $j=0$ term of the
second series on the right-hand side of the bound in
Lemma~\ref{full-short-time} below vanishes when $m=1$.) But the
part of $F_{\lambda}(t)$ due to the exchange term is $O(\lambda \|
F_{\lambda} \|_1)$ by the above estimate, and $\lambda \|
F_{\lambda} \|_1$ already tends to zero.

Knowing that the exchange effect vanishes in the mean field limit,
one might wonder whether
Theorem \ref{mean-field-theorem} is trivial.  The theorem states that
the error between the
true single-particle density and its TDHF approximation tends to
zero under certain conditions --- but perhaps this so simply because the
effect of the interaction disappears in the limits we have taken?
At least the ``direct" part of the interaction does not disappear in the limit,
for it survives as the nonlinear term in the time-dependent Hartree equation as
shown in the following corollary of Theorem \ref{mean-field-theorem}:
\begin{corollary}
\label{corollary}
Let $\psi \in L^2(\RRR^3)$ have norm one, and for $n \in \NNN$ let
$\Psi_n(x,t) \in L^2(\RRR^{3n})$ be the solution of the
Schr\"odinger equation
\begin{eqnarray}
          i \frac{\partial}{\partial t} \Psi_n(x,t)  & = &
-\frac{1}{2}\sum_{j=1}^n \Delta_{x_j} \Psi_n(x,t) \ + \
          \frac{1}{n}\sum_{i < j}V(|x_i - x_j|) \Psi_n(x,t) \nonumber \\
         \Psi_n(x,0) & = & \psi(x_1)\psi(x_2)\cdots\psi(x_n)
         \label{BEC}
\end{eqnarray}
with $V(x)$ bounded.  Let $\rho_n(x,y,t)$ denote the integral
kernel of the single-particle density operator, normalized to have
trace $1$ rather than $n$:
\begin{equation}
\label{rho}
        \rho_n(x,y,t)   \ = \ \int_{\RRR^3} \cdots \int_{\RRR^3}
\overline{\Psi_n(y,z_2,\ldots,z_n,t)}\Psi_n(x,z_2,\ldots,z_n,t)dz_2\cdots
        dz_n\ .
\end{equation}
Let $\rho_*(x,y,t)$ denote the solution of the time-dependent Hartree equation
\begin{eqnarray}
     i \frac{\partial}{\partial t} \rho_*(x,y,t) & = & -\tfrac{1}{2}(\Delta_x -
     \Delta_y)\rho_*(x,y,t) \ + \   \int_{\RRR^3}
         \big[ V(|x-z|) -  V(|y-z|) \big]
         \rho_*(z,z,t)dz \  \rho_*(x,y,t)
     \nonumber \\
     \rho_*(x,y,0) & = & \overline{\psi(y)}\psi(x) \ .
     \label{bosons}
\end{eqnarray}
Then $\rho_n(t)$ converges in trace norm to $\rho_*(t)$ at each
fixed $t \ge 0$ as $n \longrightarrow \infty$.
\end{corollary}
The proof of this corollary is given in the appendix.

\noindent {\bf Proof of Theorems~\ref{short-times} and
\ref{mean-field-theorem}}

Let $\NN_m(t)$ be as in
Proposition~\ref{BBGKY}, and let $\FF_m(t)$ satisfy the TDHF
hierarchy.  In the hypotheses Theorems~\ref{short-times} and
\ref{mean-field-theorem} we suppose that $F(0)=\NN_1(D(0))$, but
for now let us only assume that
\begin{equation}
\label{boundsOnF}
      \|F(0)\|\le 1 \qquad \hbox{and} \qquad \|F(0)\|_1 =
      \|\NN_1(D(0))\|_1.
\end{equation}
The trace norm of $\NN_1(t)$ is independent of $t$, and we shall
denote it simply by $\|\NN_1 \|_1$.
  Assuming (\ref{boundsOnF}), the
bound of Proposition~\ref{TDHFbound} is itself bounded by
\begin{equation}
\label{important-estimate}
        m(m-1)2\|V\| \ \big\|\NN_1 \big\|_1^m t.
\end{equation}
  Subtracting equations (\ref{FFm(t)}) from
the BBGKY equations of Proposition~\ref{BBGKY} and using
(\ref{important-estimate}) leads to the estimates
\begin{eqnarray*}
              \big\| \NN_m(t) - \FF_m(t) \big\|_1
          & \le &
          \big\| \NN_m(0) - \FF_m(0) \big\|_1  \ + \
           m(m-1)\|V\|
                  \Big( 2 \| \NN_1 \|_1^m +  \| \NN_m \|_1 \Big)t  \\
          & + &
          m 2 \|V\|
                 \int_0^t \big\| \NN_{m+1}(s) - \FF_{m+1}(s) \big\|_1 ds \ .
\end{eqnarray*}
Iterating this estimate $k$ times, one obtains
\begin{eqnarray}
              \big\| \NN_m(t) - \FF_m(t) \big\|_1
          & \le &
          \sum_{j=0}^{k} a_{m+j} \binom{m+j-1}{j}   C^j t^j
          \ + \
          \sum_{j=0}^{k} \frac{b_{m+j}}{j+1}
         \binom{m+j-1}{j}   C^j t^{j+1}
          \nonumber \\
          & + &
          C^n \frac{(m+k)!}{(m-1)!}
           \int_0^t \int_0^{t_1} \cdots \int_0^{t_{n}}
             \big\| \NN_{m+k+1}(s) - \FF_{m+k+1}(s) \big\|_1
         ds dt_k \cdots dt_1 \nonumber \\
\qquad \label{key-estimate}
\end{eqnarray}
with $ C = 2\|V\|$ and
\begin{eqnarray}
         a_m & = & \big\| \NN_m(0) - \FF_m(0) \big\|_1 \nonumber \\
     b_m & = & m(m-1)\|V\|
                  \Big( 2 \| \NN_1 \|_1^m  +  \| \NN_m \|_1 \Big).
\label{key-notation}
\end{eqnarray}
To make use of these estimates we need some control over the size
of the integrand in (\ref{key-estimate}). We will assume that
\begin{equation}
        \| \NN_m \|_1 \ \le \  \| \NN_1 \|_1^m
\label{moment-hypothesis}
\end{equation}
for all $m$, for this bound holds for gauge-invariant quasifree
state with finite expected particle number
  by (\ref{exactClosure}) and Proposition \ref{Prop1.5}.
  Note that (\ref{moment-hypothesis}) holds independently of time,
since the dynamics conserve particle number.
  With the bound
(\ref{moment-hypothesis}), the last term on the right hand side of
(\ref{key-estimate}) may be bounded by
\[
     2\binom{m+k}{m-1}\big( C \big\| \NN_1 \big\|_1 t\big)^{m+k+1} ,
\]
which tends to $0$ as $k$ tends to infinity if $m$ fixed and
$C\|\NN_1\|_1t < 1$.   Furthermore, assuming
(\ref{moment-hypothesis}), we can bound $b_m$ of
(\ref{key-notation}) by $ \tfrac{3}{2} C \| \NN_1 \|_1^m m(m-1)$
and establish the following lemma:
\begin{lemma}
\label{full-short-time}
  Suppose that $D$ is a density operator on
$\FFF_{\HHH}$ of the form $D=\oplus D_n$, such that
(\ref{moment-hypothesis}) holds for all $m \in \NNN$.  Let $\WW_t$
be as defined in (\ref{dynamics}) and let $\NN_m(t)$ denote
$\NN_m(\WW_t(D))$. Let $F(t)$ be the solution of a TDHF equation
(\ref{TDHF-mild}) whose initial condition $F(0)$ satisfies
(\ref{boundsOnF}), and let $\FF_m(t)$ be as in (\ref{FFm}). Then,
with $C=2\|V\|$,
\begin{eqnarray*}
             \frac{ \big\| \NN_m(t) - \FF_m(t) \big\|_1 }{\big\| \NN_1
\big\|_1^m}
                & \le &
          \sum_{j=0}^\infty
          \frac{\big\| \NN_{m+j}(0) - \FF_{m+j}(0) \big\|_1 } {\big\|
\NN_1 \big\|_1^{m+j} }
           \binom{m+j-1}{m-1}
            \big( C \| \NN_1 \|_1  t\big)^j \\
            & + &
           \big\| \NN_1 \big\|_1^{-1} \frac{3}{2}
               \sum_{j=0}^\infty (m+j-1) \binom{m+j}{m-1}
          \big( C \| \NN_1 \|_1  t\big)^{j+1}
\end{eqnarray*}
when $ C \| \NN_1 \|_1  t <1 $.
\end{lemma}

In the hypotheses of Theorems~\ref{short-times} and
\ref{mean-field-theorem}, the initial data for the exact dynamics
are assumed to be gauge-invariant quasifree states $D$ (or
$D_{\lambda}$) with finite expected particle number, and the
initial data for the corresponding TDHF equations are assumed to
be $F(0) = \NN_1(D)$ (or $F_{\lambda}(0) = \NN_1(D_{\lambda})$).
Thus, the requirements (\ref{moment-hypothesis}) and
(\ref{boundsOnF}) in Lemma~\ref{full-short-time} are satisfied
under the hypotheses of Theorems~\ref{short-times} and
\ref{mean-field-theorem}.

Theorem~\ref{short-times} follows from Lemma~\ref{full-short-time}
since $\NN_m(0) = \FF_m(0)$ for all $m$ by (\ref{exactClosure}).

To prove Theorem~\ref{mean-field-theorem} we apply
Lemma~\ref{full-short-time} to the many-particle system
(\ref{dynamics}) and the TDHF equation (\ref{TDHF-mild}) with
$\lambda V$ in place of $V$.

Statement~1 of the theorem comes easily:   $\NN_{\lambda m}(0) =
\FF_{\lambda m}(0)$ for all $m$, and Lemma~\ref{full-short-time}
implies that
\[
             \frac{ \big\| \NN_{\lambda m}(t) -
         \FF_{\lambda m}(t) \big\|_1 }{\|\NN_{\lambda m}\|_1^m}
                \  < \ \lambda C  t \ \frac{3}{2}
               \sum_{j=0}^\infty \frac{(m+j)^m}{(m-1)!}
          \big( \lambda C \|\NN_{\lambda 1}\|_1 t \big)^j \nonumber \\
\]
with $C=2\|V\|$ when $ \lambda  C \| \NN_{\lambda 1} \|_1  t < 1
$. Statement~1 follows since $\lim\limits_{\lambda\rightarrow 0}
\lambda\|\NN_{\lambda 1}\|_1 = 0$.

The proof of Statement~2 requires the fuller version of the
inequality in Lemma~\ref{full-short-time}.  Since
(\ref{moment-hypothesis}) and (\ref{boundsOnF}) are satisfied at
any time $s > 0$ if they are satisfied initially,
Lemma~\ref{full-short-time} implies that
\begin{eqnarray}
             \frac{ \big\| \NN_{\lambda m}(s+\Delta t) -
         \FF_{\lambda m}(s+\Delta t) \big\|_1 }{\|\NN_{\lambda 1}\|_1^m}
                & \le &
          \sum_{j=0}^\infty
          \frac{\big\| (\NN_{\lambda})_{m+j}(s) -
      (\FF_{\lambda})_{m+j}(s) \big\|_1 } {\|\NN_{\lambda 1}\|_1^{m+j} }
           \binom{m+j-1}{m-1}
            \big( \lambda C \|\NN_{\lambda 1}\|_1 \Delta t \big)^j \nonumber \\
            & + &
           \lambda C \Delta t \frac{3}{2}
               \sum_{j=0}^\infty (m+j-1) \binom{m+j}{m-1}
          \big( \lambda C \|\NN_{\lambda 1} \|_1 \Delta t \big)^j \nonumber \\
\label{take-limit}
\end{eqnarray}
for any $s \ge 0$ as long as $\Delta t < (\lambda C \|\NN_{\lambda
1}\|_1)^{-1}$. Since $
       u =  C \limsup\limits_{\lambda\rightarrow 0}
       \lambda \|\NN_{\lambda 1}\|_1
$ is finite by hypothesis, taking the $\limsup$ of both sides of
(\ref{take-limit}) shows that
\begin{equation}
\label{Slater-closure}
       \lim_{\lambda\rightarrow 0}  \big\| \NN_{\lambda m}(t)
       - \FF_{\lambda m}(t)  \big\|_1
       \big/
        \|\NN_{\lambda 1}\|_1^{m}
       \ = \ 0 \qquad \forall \ m \in \NNN
\end{equation}
holds at time $t=s+\Delta t$ if it holds at time $t=s$ and $\Delta
t < 1/u$. Since (\ref{Slater-closure}) holds at $t=0$, an
inductive argument proves that it holds at all times $t>0$.
  \hfill $\square$


\section{Appendix: the time-dependent Hartree equation}
Here we prove Corollary~\ref{corollary}:

The initial
condition of (\ref{BEC})
is not available to $n$ fermions in $\RRR^3$, for it is in extreme
violation of the Pauli Exclusion Principle.
   To derive (\ref{bosons}) from
our theorem about fermions we will introduce an auxilliary space
$\ell^2(\NNN)$ to allow Pauli exclusion to hold while the spatial
part of the $n$-particle wavefunction is permitted have the form
$\psi(x_1)\psi(x_2)\cdots\psi(x_n)$.  We are going to apply
Statement~2 of Theorem~\ref{mean-field-theorem} where the
single-particle Hilbert space $\HHH = L^2(\RRR^3)\otimes
\ell^2(\NNN)$.

Let $e_1,e_2,\ldots$ be an orthonormal sequence in $\ell^2(\NNN)$
and let $D_{1/n}(0)$ be the orthogonal projector on $\FFF_{\HHH}$
whose range is the span of the $n$-particle Slater determinant
\[
      A_n \left(
      (\psi\otimes e_1) \otimes (\psi\otimes e_2) \otimes
      \cdots \otimes (\psi\otimes e_n) \right).
\]
It is helpful to rearrange factors and write
\[
D_{1/n}(0) = R_n(P_{\psi^{\otimes n}} \otimes S_n)R_n^*
\ ,
\]
where $P_{\psi^{\otimes n}}$ denotes projection onto the span of
$\psi^{\otimes n}$, and $S_n$ denotes the rank-one orthogonal
projector on $\ell^2(\NNN)^{\otimes n}$ whose image is the span of
the Slater determinant formed from $e_1,\ldots,e_n$, and $R_n$ is
the unitary transformation from $L^2(\RRR^3)^{\otimes n} \otimes
\ell^2(\NNN)^{\otimes n}$ to
$\left(L^2(\RRR^3)\otimes\ell^2(\NNN)\right)^{\otimes n}$  that
rearranges the factors of any simple tensor thus:
\[
       R_n((f_1 \otimes f_2 \otimes \cdots \otimes f_n ) \otimes
(s_1\otimes s_2 \otimes \cdots \otimes s_n))
  \ = \ (f_1 \otimes s_1) \otimes \cdots \otimes (f_n \otimes s_n)
  \ .
\]
If $\Psi_n(t)$ is the solution of (\ref{BEC}) then $P_{\Psi_n(t)}$
--- the projector whose range is the span of $\Psi_n(t)$ --- satisfies
the von Neumann equation
\begin{eqnarray}
    i \frac{d}{dt} P_{\Psi_n(t)} & = & -\frac{1}{2}\sum_{j=1}^n
[\Delta_{x_j},\ P_{\Psi_n(t)}]
    \ + \ \frac{1}{n}\sum_{i < j}[M_{Vij},\ P_{\Psi_n(t)}]
    \nonumber \\
       P_{\Psi_n(t)} & = & P_{\psi^{\otimes n}}
\label{originalLiouville}
\end{eqnarray}
wherein $M_{Vij}$ denotes the multiplication operator $
      M_{Vij} \phi(x) \ = \ V(|x_i-x_j|)\phi(x).
$ Thus
\begin{eqnarray}
    i \frac{d}{dt}  R_n(P_{\Psi_n(t)} \otimes S_n)R_n^*     & = &
    -\frac{1}{2}\sum_{j=1}^n R_n([\Delta_{x_j},\ P_{\Psi_n(t)}]
\otimes S_n)R_n^*
    \ + \  \frac{1}{n}\sum_{i < j}
    R_n([M_{Vij},\ P_{\Psi_n(t)}] \otimes S_n)R_n^*
   \nonumber \\
     R_n(P_{\Psi_n(0)} \otimes S_n)R_n^*  & = & D_{1/n}(0) \ .
\label{fromOriginalLiouville}
\end{eqnarray}
  Now we define
\[
      D_{1/n}(t) \ = \ R_n(P_{\Psi_n(t)} \otimes S_n)R_n^*
\]
for all $t \ge 0$, and we use the same notation to denote the
extension of $D_{1/n}(t)$ to a density operator defined on all of
$\FFF_{\HHH}$.  From (\ref{fromOriginalLiouville}) it may be seen
that $D_{1/n}(t)$ is the solution of the von Neumann
equation
\begin{eqnarray*}
    i \frac{d}{dt} D_{1/n}(t) & = & [H_{1/n},D_{1/n}(t)]  \\
     D_{1/n}(0) & = & R_n(P_{\psi^{\otimes n}} \otimes S_n)R_n^*
\end{eqnarray*}
where $H_{1/n}$ is the Hamiltonian (\ref{LambdaHam}) on
$\FFF_{\HHH}$ with $L = -\tfrac{1}{2}\Delta\otimes
I_{\ell^2(\NNN)}$ and $V$ denoting the multiplication operator
\[
             ( V\phi )(x,j,y,k) \ = \ V(|x-y|)\phi(x,j,y,k)
\]
on $\left(L^2(\RRR^3)\otimes\ell^2(\NNN)\right)^{\otimes 2}$ --- a
slight abuse of notation.

Set $\lambda=1/n$.  According to Statement~2 of
Theorem~\ref{mean-field-theorem}
\[
       \lim\limits_{\lambda \rightarrow 0}
       \big\| \NN_{\lambda 1}(t) - F_{1/n}(t)
       \big\|_1 \big/ \|\NN_{\lambda 1}\|_1
       \ = \ 0
\]
for all $t>0$, where $\NN_{\lambda 1}(t) = \NN_1(D_{1/n}(t))$ and
$F_{1/n}$ satisfies the TDHF equation (\ref{TDHF-lambda}).  One
may verify that the single-particle density operator
$\NN_1(D_{1/n}(t))$ equals $\rho_n(t) \otimes P_n$ where $P_n$
denotes the orthogonal projector onto the span of
$\{e_1,\ldots,e_n\} \subset \ell^2(\NNN)$ and $\rho_n(t)$ is as
defined in (\ref{rho}).  On the other hand, we claim that
$F_{1/n}(t) = \xi_{\lambda}(t) \otimes P_n$ where $\xi_{\lambda}(t)$
denotes the operator on $L^2(\RRR^3)$ whose integral kernel
satisfies
\begin{eqnarray}
     i \frac{\partial}{\partial t} \xi_{\lambda}(x,y,t) & = &
-\tfrac{1}{2}(\Delta_x -
     \Delta_y)\xi_{\lambda}(x,y,t) \nonumber \\
     &  & + \ \int_{\RRR^3} \big( V(|x-z|) -  V(|y-z|) \big)
\xi_{\lambda}(z,z,t)dz \ \xi_{\lambda}(x,y,t)
     \nonumber \\
    &  & - \  \frac{1}{n} \int_{\RRR^3} \big(
         V(|x-z|) - V(|y-z|)\big)
             \xi_{\lambda}(x,z,t) \xi_{\lambda}(z,y,t) dz
     \nonumber \\
     \xi_{\lambda}(x,y,0) & = & \overline{\psi(y)}\psi(x) \ = \ \rho_n(x,y,0)\ .
     \label{xi}
\end{eqnarray}
Accepting this claim for now, and noting that
$\|\NN_1(D_{1/n}(t))\|_1$ is always equal to $n$, we find that
\begin{equation}
       \lim\limits_{n \rightarrow \infty}
       \big\| \rho_n(t) - \xi_{\lambda}(t)
       \big\|_1      \ = \ 0
\label{almostThere}
\end{equation}
for all $t>0$, since $\big\| \NN_{\lambda 1}(t) -
F_{1/n}(t)\big\|_1 =
  \big\|(\rho_n(t)- \xi_{\lambda}(t))\otimes P_n \big\|_1
= n \ \big\| \rho_n(t)-\xi_{\lambda}(t)\big\|_1$.
   Equation (\ref{xi}) is a small perturbation of (\ref{bosons}) when
$n$ is large, and
we may verify that $\xi_{\lambda}(t)$ converges to $\rho_*(t)$ by
applying Gronwall's inequality to
\begin{eqnarray}
   i \frac{d}{dt} ( \rho_* - \xi_{\lambda})
   & = & [L,\ \rho_* - \xi_{\lambda}] \ + \ [V,\ (\rho_* -
\xi_{\lambda})\otimes \rho_*]_{:1}
             \nonumber \\
    & + &  [V,\ \xi_{\lambda} \otimes (\rho_* - \xi_{\lambda})]_{:1}
    \ + \ \frac{1}{n}[V,\ (\xi_{\lambda}\otimes \xi_{\lambda})U]_{:1} \ .
\label{reallyAlmostThere}
\end{eqnarray}
First, we get rid of the term $[L,\ \rho_* - \xi_{\lambda}]$ by
passing to the ``interaction picture" and rewriting
(\ref{reallyAlmostThere}) as an equation for $\exp(itL)(\rho_* -
\xi_{\lambda})\exp(-itL)$.  Upon integrating and taking the trace
norm one obtains
\[
     \big\| \rho_*(t)  - \xi_{\lambda}(t) \big\|_1 \ \le \  4\|V\|
\int_0^t  \big\| \rho_*(s)  - \xi_{\lambda}(s) \big\|_1 ds
                \ + \ \frac{2}{n}\|V\|  \ .
\]
Gronwall's inequality implies that $ \big\| \xi_{\lambda}(t) -
\rho_*(t) \big\|_1  \ = \ O(1/n) $ as $n \longrightarrow \infty$
for fixed $t$, which implies with (\ref{almostThere}) that $
\big\| \rho_n(t) - \rho_*(t) \big\|_1 \longrightarrow 0$, as
asserted in the corollary.

Finally, we verify that $\xi_{\lambda}(t) \otimes P_n$ satisfies
(\ref{TDHF-lambda}) when $\xi_{\lambda}(t)$ satisfies (\ref{xi}):
\begin{eqnarray}
          i \frac{d}{dt} F_{1/n} & = &
          i \frac{d}{dt} \xi_{\lambda}\otimes P_n
     \ = \ \left([-\tfrac{1}{2}\Delta,\xi_{\lambda}] + \big[M_V,\
\xi_{\lambda}^{\otimes 2}(I-\tfrac{1}{n}U)\big]_{:1}\right)
      \otimes P_n  \nonumber \\
& = &
          [L,F_{1/n}] \ + \
      \frac{1}{n} \hbox{Trace}_{3,4}\big[V,\ (\xi_{\lambda}\otimes
P_n)^{\otimes 2}\big]
     \nonumber  \\
& & - \ \frac{1}{n} \hbox{Trace}_{3,4} \big[V,\
(\xi_{\lambda}\otimes P_n)^{\otimes 2}
  U_{(12)(34)}\big]
\label{FullTDHF}
\end{eqnarray}
where $\hbox{Trace}_{3,4}$ denotes the partial trace over the
third and fourth factors of the tensor product, and
$U_{(13)(24)}$ denotes the unitary operator on $(L^2(\RRR)\otimes \ell^2(\NNN))^{\otimes 2}$ that exchanges the first and
third factors simple tensor
products as it exchanges their second and fourth factors. The second term on the right-hand side of
(\ref{FullTDHF}) is multiplied by $1/n$ to compensate for the
$\hbox{Trace}(P_n)=n$ due to the partial trace over the fourth
factor.  But (\ref{FullTDHF}) is the differential form of
(\ref{TDHF-lambda}) for $\lambda = 1/n$. $\hfill \square$


\bigskip

{\bf Acknowledgement.} {\em This research was supported by  by the
European network HYKE (contract HPRN-CT-2002-00282), the
French-Austrian ``Amadeus" program (\"OAD 19/2003)) and by the
Austrian START project ``Nonlinear Schr\"odinger and quantum
Boltzmann equations" of N.J.M. (contract Y-137-Tec).
F.G. acknowledges support from the French IUF.

We also express our gratitude to Xavier Blanc, Eric Canc\`es and
Claude Le Bris for numerous valuable discussions.}


\end{document}